\documentclass[aip,jcp,numerical,reprint]{revtex4-1}
\usepackage{graphicx}  
\usepackage{threeparttable} 
\usepackage{siunitx} 
\usepackage{bm,amssymb,amsmath} 
\usepackage[bookmarks]{hyperref} 
\usepackage[svgnames]{xcolor} 
\definecolor{Blue}{RGB}{0 0 128}
\hypersetup{ 
	colorlinks= true,
    allcolors = Blue
}
\usepackage{tikz}
\usetikzlibrary{calc} 
\usetikzlibrary{arrows,decorations.markings} 
\usepackage{color,soul} 
\begin{document}

\title{How Large are Nonadiabatic Effects in Atomic and Diatomic Systems?}
\author{Yubo Yang}
\affiliation{Department of Physics, University of Illinois, Urbana, Illinois 61801 USA}
\author{Ilkka Kyl\"{a}np\"{a}\"{a}}
\affiliation{Department of Physics, Tampere University of Technology, P.O.~Box 692, FI-33101 Tampere, Finland}
\affiliation{Department of Physics, University of Illinois, Urbana, Illinois 61801 USA}
\author{Norm M. Tubman}
\affiliation{Department of Physics, University of Illinois, Urbana, Illinois 61801 USA}
\author{Jaron T. Krogel}
\affiliation{Materials Science \& Technology Division, Oak Ridge National Laboratory, Oak Ridge, TN 37831}
\author{Sharon Hammes-Schiffer}
\affiliation{Department of Chemistry, University of Illinois, Urbana, Illinois 61801 USA}
\author{David M. Ceperley}
\affiliation{Department of Physics, University of Illinois, Urbana, Illinois 61801 USA}
\date{\today}

\begin{abstract}
With recent developments in simulating nonadiabatic systems to high accuracy, it has become possible to determine how much energy is attributed to nuclear quantum effects beyond zero-point energy. In this work we calculate the non-relativistic ground-state energies of atomic and molecular systems without the Born-Oppenheimer approximation. For this purpose we utilize the fixed-node diffusion Monte Carlo method, in which the nodes depend on both the electronic and ionic positions. We report ground-state energies for all systems studied, ionization energies for the first-row atoms and atomization energies for the first-row hydrides. We find the ionization energies of the atoms to be nearly independent of the Born-Oppenheimer approximation, within the accuracy of our results. The atomization energies of molecular systems, however, show small effects of the nonadiabatic coupling between electrons and nuclei.
\end{abstract}
\maketitle

\section{Introduction}
There have been several recent discoveries~\cite{Tubman_ECG,cederbaum1,gross2014,boent,Martinez_Review} suggesting that quantum wave functions, which include both electronic and ionic degrees of freedom, have many interesting properties that have yet to be explored.  This includes the development of equations that exactly factorize a wave function into electronic and ionic components,~\cite{cederbaum1,cederbaum12} the disappearance of conical intersections in wave functions of model systems,~\cite{gross2014} and the use of quantum entanglement to study electronic and ionic density matrices.~\cite{boent} Extending such studies to realistic systems is of broad interest and will considerably expand our understanding of electron-ion systems. However, treatment of \textit{ab initio} electron-ion systems is challenging, and applications have thus been limited. The most accurate simulations of electron-ion wave functions are generally done with very specialized wave functions, which are limited to rather small systems.~\cite{mitroy2013} Methods are also being developed to treat larger systems with different regimes of validity.~\cite{Sharon_NEO-HF,Sharon_XCNEO-HF1,Sharon_XCNEO-HF2,Sharon_XCNEO-HF,Kurt_XCNEO-HF,Kurt_XCNEO-HF1,Sharon_NEO-DFT,Sharon_NEO-DFT2,Sharon_NEO-DFT3,Gross_NEO-DFT,Gross_NEO-DFT1,Ilkka_Path,Ilkka_Path1,Ilkka_Path2}

As a framework to address these problems in general realistic systems, we recently demonstrated that quantum Monte Carlo (QMC) can be combined with quantum chemistry techniques to generate electron-ion wave functions.~\cite{Tubman_ECG} We treated realistic molecular systems and demonstrated that our method can be scaled to larger systems than previously considered while maintaining a highly accurate wave function. In the following we extend our previous work by considering the simulation of a larger set of atoms and molecules.  We calculate ionization energies and atomization energies that can be directly compared with previous results for benchmarking purposes.

\section{Method}
\subsection{Fixed-Node Diffusion Monte Carlo (FN-DMC)}
Diffusion Monte Carlo~\cite{Anderson_DMC,lester1,Stuart_Review,Needs_Review,Needs_Old_Review,QMC_Review} is a projector method that evolves a trial wave function in imaginary time and projects out the ground-state wave function. For practical simulations of fermions, the fixed-node approximation is introduced, which depends only on the set of electronic positions where a trial wave function is equal to zero.  This approximation is different than approximations typically used in quantum chemistry calculations, and in this work we demonstrate that we can generate high-quality nodal surfaces for a range of systems that include full electron-ion wave functions. 

If the trial wave function has the same nodal surface as the exact ground-state wave function, FN-DMC will obtain the exact ground-state energy.  Approximate nodal surfaces can be generated through optimization of the full wave function. Such approximate nodal surfaces have been tested and validated on a wide range of systems, and consistently provide an excellent approximation of the exact ground-state energy, comparable to the state of the art in \textit{ab initio} simulations.~\cite{grossman1} In addition, the energies generated with FN-DMC are variational with respect to the ground-state energy.

In all but a handful of previous QMC simulations,~\cite{Ceperley_1987,Natoli_1993,Natoli_1995,Chen_1995,Coldwell_H2_2008,Gabriele_H2_2004,Sandro_finiteT-noBO_2012} calculations are performed with nuclei "clamped" to their equilibrium positions. However, such an assumption is not fundamentally required by FN-DMC. 

\subsection{Electronic Wave Function and Optimization}

There are several different approaches for generating electronic wave functions for a FN-DMC calculation.~\cite{Umrigar_Alleviation,Toulouse_Bench,Brown_Bench,Seth_Bench} Recent advances~\cite{Nightingale_Linear,Umrigar_Linear,Brown_Bench} have made it possible to simultaneously optimize thousands of wave function parameters using variational Monte Carlo with clamped nuclei. We use an initial guess for the wave function that is generated from complete active space self-consistent-field (CASSCF)~\cite{Chaban_MCSCF,Szabo} calculations using the quantum chemistry package GAMESS-US.~\cite{GAMESS} The optimized orbitals are then used in a configuration interaction singles and doubles (CISD) calculation to generate a series of configuration state functions (CSFs).~\cite{Pauncz_CSF} For the small systems Li$^+$, Be$^+$, LiH and BeH, a CASSCF calculation with a large active space is used in place of CISD. The multi-CSF expansion of the wave function can be expressed in the following form,
\begin{align}
\Psi_{\text{CISD}}(\vec{r};\vec{R}_o)=\sum\limits_{i=1}^{N_{\text{CSF}}}\alpha_i\phi_i(\vec{r};\vec{R}_o), \label{eq:psi_gms}
\end{align}
where $\vec{r}$ refers to the spatial coordinates of all the electrons and $\vec{R}_o$ refers to the equilibrium positions of all the ions. $\phi_i(\vec{r})$ and $\vec{\alpha}=\{\alpha_1,\alpha_2,\dots\}$ are the CSFs and CI coefficients generated from CISD. The cc-pV5Z basis~\cite{dunning} is used for the atomic systems and the Roos Augmented Triple Zeta ANO basis~\cite{roos} is used for the molecular systems except for the smallest system LiH, where the cc-pV5Z basis is used.

After the multi-CSF expansion is generated, we impose the electron-nucleus cusp condition on each molecular orbital~\cite{cusp} and add a Jastrow factor to the wave function to include electron correlation.~\cite{Kato} Our Jastrow factor contains electron-electron, electron-nucleus and electron-electron-nucleus terms. The full electronic wave function used in FN-DMC is,
\begin{align}
\psi_e(\vec{r};\vec{R})=e^{J(\vec{r},\vec{R},\vec{\beta})}\Psi_{\text{CISD}}(\vec{r};\vec{R})\label{eq:psie}.
\end{align}
We optimize the CSF and Jastrow coefficients, $\vec{\alpha}$ and $\vec{\beta}$, respectively, simultaneously with QMCPACK.~\cite{QMCPACK_Kim,QMCPACK_Esler} Optimization is performed with the ions clamped to their equilibrium positions $\vec{R}_o$. The equilibrium geometries for BeH and BH are chosen to be the ECG-optimized distances for comparison with the ECG (explicitly correlated Gaussian) method, and the geometries for the rest of the hydrides are taken from experimental data.~\cite{CCCBDB} We use 3.015 a.u. as the equilibrium inter-nuclei distance for LiH, as this geometry is found to provide a lower clamped-nuclei ground-state energy than the ECG optimized distance of 3.061 a.u.. We include all CSFs with coefficients larger than a specific cutoff $\epsilon$ to lend reasonable flexibility to the wave function during optimization. We include as many CSFs as possible to maximize the flexibility of the wave function. However, the inclusion of too many CSFs with small expansion coefficients can introduce noise as they require a large number of samples in the optimization step to be optimized. We have chosen $\epsilon$ to restrict the number of CSFs in the wave function to be $\sim$1000 in all systems studied. Optimization is performed with the linear method~\cite{Umrigar_Linear} with roughly $10^6$ statistically independent samples.


\begin{figure}[h]
\includegraphics[width=9cm]{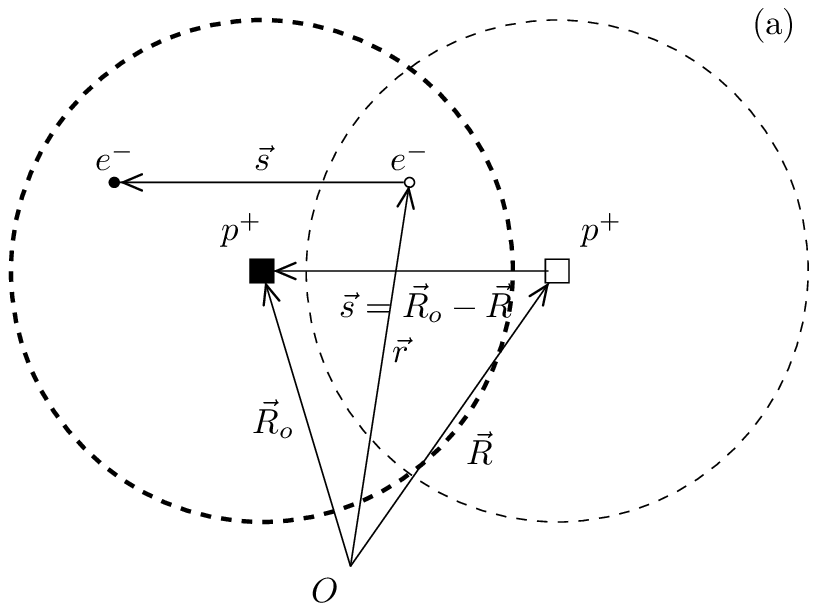}
\includegraphics[width=9cm]{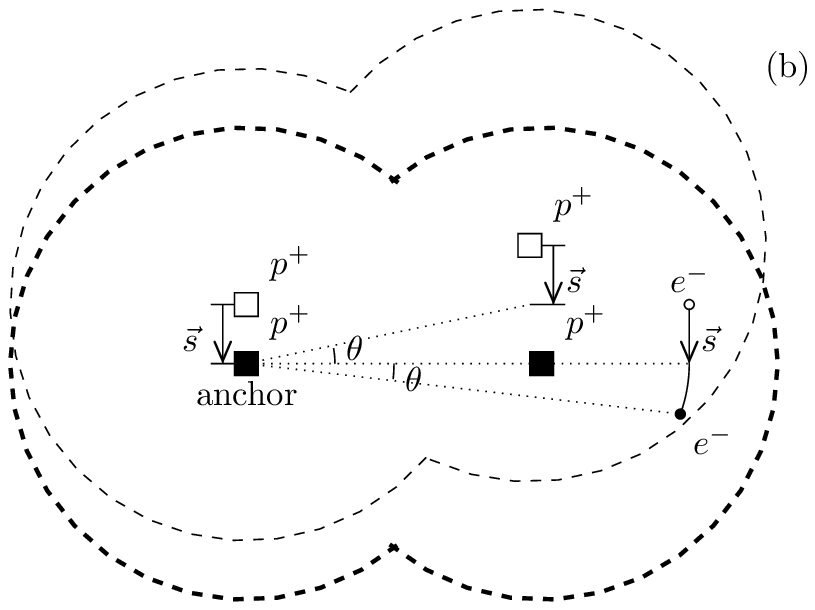}
\caption{Dragged-node approach for simulation of atomic and molecular systems in QMC. {\bf (a)} For atomic systems we can consider the entire wave function shifting with the ion. This process can be visualized by following a contour of the wave function. The thick dashed circle represents a contour of the electronic wave function when the proton is at its reference position $\vec{R}_o$, and the thin dashed circle represents the same contour when the proton has moved to a new position $\vec{R}$. To evaluate the ion-dependent electronic wave function $\bar{\psi}_e(\vec{r},\vec{R})$, we simply map the electron to its proper place in the reference wave function $\psi_e(\vec{r};\vec{R}_o)$. That is, $\bar{\psi}_e(\vec{r},\vec{R})=\bar{\psi}_e(\vec{r}+\vec{s},\vec{R}_o)=\psi_e(\vec{r}+\vec{s};\vec{R}_o)$ where $\vec{s}$ is the shift required to put the proton back to its reference position. {\bf (b)} For H$_2^+$, we pick one of the protons as an ``anchor'' and approximate the new wave function by dragging the reference wave function with the ``anchor'' proton. We also rotate the wave function to align its axis of symmetry with the orientation of the two protons. \label{fig:drag}}
\end{figure}

\subsection{Electron-Ion Wave Function}

Once a satisfactory electronic wave function has been obtained, we construct the electron-ion wave function using the ansatz,
\begin{align}
\Psi_{eI}(\vec{r},\vec{R})=\psi_I(\vec{R})\bar{\psi}_e(\vec{r},\vec{R}), \label{eq:psi}
\end{align}
where $\vec{R}$ denotes the spatial coordinates of all ions and $\bar{\psi}_e(\vec{r},\vec{R})$ is an ion-dependent electronic wave function adapted from the clamped-nuclei wave function $\psi_e(\vec{r};\vec{R}_o)$ through basis set dependence. Due to the localization of Gaussian basis sets around nuclei, as used in quantum chemistry calculations, the nodes of $\bar{\psi}_e$ change based on the ionic positions, which we have previously called the dragged-node approximation.~\cite{Tubman_ECG}
Although there are approaches for going beyond the dragged-node approximation, it was demonstrated to be highly accurate over a range of molecules in previous work.~\cite{Tubman_ECG} For the systems considered here, we can impose various symmetries of the Hamiltonian onto the wave function that arise from the relative motion of the ions. In Fig.~\ref{fig:drag} we demonstrate this approach for the simple cases of a hydrogen atom and an H$_2^+$ molecular ion. This approach can be generalized for use in larger systems or even applied to parts of a bigger system, e.g., treating light ions as quantum particles and heavy ions as "clamped".

The term $\psi_I$ consists of simple products of Gaussian wave functions over each pair of nuclei,
\begin{align}
\psi_I(\vec{R})\propto \prod\limits_{i,j>i}e^{-a_{ij}(\vert \vec{R}_i-\vec{R}_j\vert-b_{ij})^2},
\label{wfs_ions}
\end{align}
where $a_{ij}$ is a coefficient that is optimized and $b_{ij}$ are taken to be the equilibrium distances between the nuclei. Since $\psi_I$ is nodeless, the choice of the variational parameters $a_{ij}$ and $b_{ij}$ does not affect the converged FN-DMC energy. FN-DMC is then performed with the fully optimized electron-ion wave function. We perform timestep extrapolation for all of the tested systems. At least four timesteps from $0.005~\text{Ha}^{-1}$ to $0.0005~\text{Ha}^{-1}$ are used for all systems studied in the clamped-nuclei FN-DMC calculation, and at least three timesteps from $0.005~\text{Ha}^{-1}$ to $0.0001~\text{Ha}^{-1}$ are used in the nonadiabatic FN-DMC calculation.

Using definitions from Ref.~\cite{Cederbaum_Review}, the adiabatic approximation will refer to the complete neglect of the nonadiabatic coupling matrix when the Schr$\ddot{\text{o}}$dinger equation is expressed in the basis of eigenstates of the electronic Hamiltonian. In this context, the nonadiabatic contribution to an eigenvalue of the electronic Hamiltonian can be partitioned into two parts: the diagonal Born-Oppenheimer correction (DBOC), which only involves the single electronic state of interest, and the remaining corrections arising from terms that involve excited eigenstates of the electronic Hamiltonian. The DBOC discussed in this work is the expectation value of the nuclear kinetic energy operator for the ground adiabatic electronic state. We define the clamped-nuclei ground-state energy $E_c$ as the lowest eigenvalue of the electronic Hamiltonian and the nonadiabatic ground-state energy $E_n$ as the lowest eigenvalue of the full molecular Hamiltonian that includes the nuclear kinetic energy. The zero-point energy (ZPE) for a diatomic molecule is the energy of the ground vibrational state of the one-dimensional vibrational mode. Note that the ZPE of the nuclei is part of the difference $E_n-E_c$. The ZPE is not considered to be nonadiabatic, but its contribution is included in the full molecular Hamiltonian.

\section{Results and Discussion}
\begin{table*}[t!]
\setlength{\extrarowheight}{1pt}
\begin{threeparttable}

\caption{Ground-state energies for atoms and ions and the ionization energies for the atoms:  fixed-node DMC results of this work (FN-DMC) for atoms and ions with and without the Born-Oppenheimer approximation. The rows marked with bold \textbf{FN-DMC} are our nonadiabatic results. The ionization potentials (IPs) are reported in the last section of the table. Energies are given in units of Hartree. For the highly accurate Hylleraas and ECG results, up to 8 digits are reported in the table. \label{tab:ionization}}
\begin{tabular}
{
 l
 S[table-format=1.6]
 S[table-format=5.6]
 S[table-format=5.6]
 S[table-format=4.6]
 S[table-format=4.6]
 S[table-format=4.6]
 S[table-format=4.6]
 S[table-format=4.6]
}

\hline\hline
\multicolumn{1}{c}{Atom} & 
\multicolumn{1}{c}{Li$(^2$S)} &
\multicolumn{1}{c}{Be$(^1$S)} &
\multicolumn{1}{c}{B$(^2$P)} &
\multicolumn{1}{c}{C$(^3$P)} &
\multicolumn{1}{c}{N$(^4$S)} &
\multicolumn{1}{c}{O$(^3$P)} &
\multicolumn{1}{c}{F$(^2$P)} \\ 
\hline
\multicolumn{1}{c}{} & 
\multicolumn{1}{c}{} &
\multicolumn{1}{c}{} &
\multicolumn{1}{c}{clamped-ion} &
\multicolumn{1}{c}{} &
\multicolumn{1}{c}{} &
\multicolumn{1}{c}{} &
\multicolumn{1}{c}{} \\
FN-DMC & \text{-}7.478057(5) & \text{-}14.66731(1) & \text{-}24.65374(2) & \text{-}37.84448(2) & \text{-}54.58851(6) & \text{-}75.0658(2) & \text{-}99.73177(6) \\
Seth DMC \cite{Seth_Bench} & \text{-}7.478067(5) & \text{-}14.667306(7) & \text{-}24.65379(3) & \text{-}37.84446(6) & \text{-}54.58867(8) & \text{-}75.0654(1) & \text{-}99.7318(1) \\
$E_{\text{ref}}$\tnote{a} &  \text{-}7.4780603 & \text{-}14.667356 & \text{-}24.653866 & \text{-}37.8450 & \text{-}54.5892 & \text{-}75.0673 & \text{-}99.7339 \\
\multicolumn{1}{c}{} & 
\multicolumn{1}{c}{} &
\multicolumn{1}{c}{} &
\multicolumn{1}{c}{nonadiabatic} &
\multicolumn{1}{c}{} &
\multicolumn{1}{c}{} &
\multicolumn{1}{c}{} &
\multicolumn{1}{c}{} \\
\textbf{FN-DMC} & \text{-}7.47742(2) & \text{-}14.66643(3) & \text{-}24.65252(4) & \text{-}37.84273(4) & \text{-}54.58641(5) & \text{-}75.06313(6) & \text{-}99.7293(1) \\
ECG \tnote{b} & \text{-}7.4774519 & \text{-}14.666435 & \text{-}24.652624 & \text{-}37.841621 & N/A & N/A & N/A \\
\hline

\multicolumn{1}{c}{Ion} & 
\multicolumn{1}{c}{$\text{Li}^+(^1$S)} &
\multicolumn{1}{c}{$\text{Be}^+(^2$S)} &
\multicolumn{1}{c}{$\text{B}^+(^1$S)} &
\multicolumn{1}{c}{$\text{C}^+(^2$P)} &
\multicolumn{1}{c}{$\text{N}^+(^4$S)} &
\multicolumn{1}{c}{$\text{O}^+(^3$P)} &
\multicolumn{1}{c}{$\text{F}^+(^2$P)} \\ 
\hline
\multicolumn{1}{c}{} & 
\multicolumn{1}{c}{} &
\multicolumn{1}{c}{} &
\multicolumn{1}{c}{clamped-ion} &
\multicolumn{1}{c}{} &
\multicolumn{1}{c}{} &
\multicolumn{1}{c}{} &
\multicolumn{1}{c}{} \\
FN-DMC & \text{-}7.27989(2) & \text{-}14.324749(7) & \text{-}24.34883(1) & \text{-}37.43071(2) & \text{-}54.05371(5) & \text{-}74.56597(6) & \text{-}99.0909(1) \\
Seth DMC \cite{Seth_Bench} & \text{-}7.279914(3) & \text{-}14.324761(3) & \text{-}24.34887(2) & \text{-}37.43073(4) & \text{-}54.05383(7) & \text{-}74.56662(7) & \text{-}99.0911(2) \\
$E_{\text{ref}}$\tnote{c} & \text{-}7.2799134 & \text{-}14.324763 & \text{-}24.348884 & \text{-}37.430880 & \text{-}54.0546 & \text{-}74.5668 & \text{-}99.0928 \\ 
\multicolumn{1}{c}{} &
\multicolumn{1}{c}{} &
\multicolumn{1}{c}{} &
\multicolumn{1}{c}{nonadiabatic} &
\multicolumn{1}{c}{} &
\multicolumn{1}{c}{} &
\multicolumn{1}{c}{} &
\multicolumn{1}{c}{} \\
\textbf{FN-DMC} & \text{-}7.27931(4) & \text{-}14.32387(2) & \text{-}24.34758(3) & \text{-}37.42899(6) & \text{-}54.05165(4) & \text{-}74.5634(1) & \text{-}99.0885(1) \\
ECG \tnote{d} & N/A &  \text{-}14.323863 &  \text{-}24.347641 &  \text{-}37.429169 & N/A & N/A & N/A \\
\hline
\multicolumn{1}{c}{} & 
\multicolumn{1}{c}{} &
\multicolumn{1}{c}{} &
\multicolumn{1}{c}{clamped-ion} &
\multicolumn{1}{c}{} &
\multicolumn{1}{c}{} &
\multicolumn{1}{c}{} &
\multicolumn{1}{c}{} \\
IP (FN-DMC) & 0.19817(2) & 0.34256(1) & 0.30490(2) & 0.41377(3) & 0.53479(8) & 0.4998(2) & 0.6409(1) \\
\multicolumn{1}{c}{} & 
\multicolumn{1}{c}{} &
\multicolumn{1}{c}{} &
\multicolumn{1}{c}{nonadiabatic} &
\multicolumn{1}{c}{} &
\multicolumn{1}{c}{} &
\multicolumn{1}{c}{} &
\multicolumn{1}{c}{} \\
IP (\textbf{FN-DMC}) & 0.19811(4) & 0.34257(4) & 0.30494(5) & 0.41374(7) & 0.53476(7) & 0.4998(1) & 0.6408(1) \\
IP (Ref.)\tnote{e} & 0.198130 & 0.342572 & 0.304980 & 0.414014 & 0.534775 & 0.500452 & 0.640946 \\
\hline\hline
\end{tabular}

\begin{tablenotes}
\item[a] For the atomic references, we use the Hylleraas result for Li,~\cite{Wang_Li} and ECG results for Be~\cite{Stanke_Be} and B.~\cite{Bubin_B} Ref.~\cite{Davidson_Atoms} is used for C,N,O and F where the ground-state energies are taken from Table XI.
\item[b] We use nonadiabatic ECG results as the reference for Li,~\cite{Stanke_Li} Be~\cite{Bubin_BeH_noBO} and B~\cite{Bubin_B}, which are converged to the true ground-state to well within 0.1 mHa. The result for C,~\cite{Bubin_C} however, may have error on the order of 1 mHa.
\item[c] For the ionic references, we use the ICI result for $\text{Li}^+$,~\cite{Nakashima_Li+} Hylleraas result for $\text{Be}^+$~\cite{Puchalski_Be+} and ECG results for $\text{B}^+$~\cite{Bubin_B+} and $\text{C}^+$.~\cite{Bubin_C+,mitroy2013} Ref.~\cite{Davidson_Atoms} is used for $\text{N}^+$,$\text{O}^+$,$\text{F}^+$.
\item[d] ECG references only exist for $\text{Be}^+$,~\cite{Bubin_BeH_noBO} $\text{B}^+$~\cite{Bubin_B+} and $\text{C}^+$.~\cite{Bubin_C+}
\item[e] Spin-orbit coupling and relativistic corrections~\cite{Klopper_IP} are removed from experimental data~\cite{NIST_Atoms} for comparison.
\end{tablenotes}

\end{threeparttable}
\end{table*}

\subsection{Atoms and Ions}

To assess the quality of our results for atoms and ions~\cite{masses}, we compare to previous results from highly accurate simulations, as presented in Table \ref{tab:ionization}. For the clamped-ion results, QMC~\cite{Brown_Bench,Toulouse_Bench,Seth_Bench,Morale_Bench,Rappe_Bench} and quantum chemistry benchmarks are available for comparison. To illustrate the high-quality QMC techniques used in this work, we compare our clamped-ion atomic results with a recent QMC benchmark study.~\cite{Seth_Bench} The ground-state FN-DMC energies consistently agree across all systems studied (except for O$^{+}$) within 0.1 mHa. This shows that similar nodes can be obtained with different forms of the wave function. In particular, our large ($\sim$ 1000 CSF) multi-determinant expansions can be compared with the approach used by Seth {\it et al.},~\cite{Seth_Bench} which relies on moderately-sized multi-determinant expansions ($\sim$ 100 CSF) with a backflow transformation. For certain atoms we can compare to more accurate simulation techniques. For C$^+$ as well as the neutral and ionized Li, Be and B, highly accurate ECG calculations that are all converged well beyond 0.1 mHa to the true ground-state energy are available. The convergence is corroborated by results from the Hylleraas method for Li~\cite{Wang_Li} and Be$^+$.~\cite{Puchalski_Be+} In Table \ref{tab:ionization} we have used the lowest variational results as our references for these systems, as the convergence is such that the accuracy is higher than other current theoretical or experimental estimates. 

\begin{figure}[h]
\centering
\includegraphics[scale=.4]{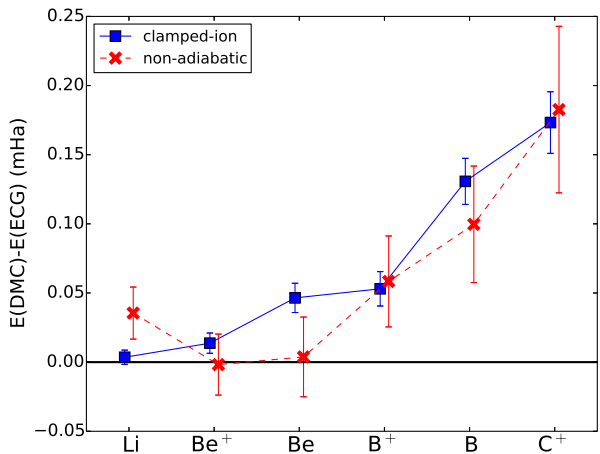}
\caption{FN-DMC ground state energies for $\text{Be}^+$, Be, $\text{B}^+$, B, $\text{C}^+$ relative to ECG references~\cite{Stanke_Be,Puchalski_Be+,Bubin_BeH_noBO,Bubin_B,Bubin_B+,Bubin_C+} for either clamped-ion or nonadiabatic calculations. These relative energies provide an estimate for the fixed-node error in the electronic and electron-ion wave functions, respectively.\label{fig:atom-ECG}}
\end{figure}

All of our clamped-ion results agree within 0.2 mHa of the ECG references, as shown in Figure \ref{fig:atom-ECG}. The error bars for the reference ECG results are absorbed into the DMC error bars for clarity, although the ECG error bars are orders of magnitude smaller compared to the DMC error bars. While ECG results exist for C and N, they are not well converged and are not suitable references.~\cite{Bubin_C,Sharkey_N} The benchmark results in Ref.~\cite{Davidson_Atoms} are a standard for atomic energies, and we report them as our references in Table \ref{tab:ionization} for the larger atoms. However, these benchmark results are not consistently accurate to 0.1 mHa. For instance, if we use the ECG results for $\text{C}^+$ with the most accurate ionization reference energy, then we find a reference energy for the C atom of -37.84489 Ha, which is 0.1 mHa higher than that reported in Ref.~\cite{Davidson_Atoms}. 
The systems with the most error are O and F, for which other QMC studies seem to experience similar difficulties.~\cite{Seth_Bench,Booth_FCIQMC,Brown_Bench,Shiwei_AFQMC} We note that for some of these systems it may be possible to absorb the sign problem and increase the accuracy further in future studies.~\cite{Tubman_Release,Tubman_ACS} 

It is more difficult to find accurate references for the nonadiabatic results. We provide the first nonadiabatic QMC benchmarks for the first-row atoms. There are six ECG calculations of nonadiabatic ground-state energies that are reportedly converged beyond 0.1 mHa, which we use as references. Our reported nonadiabatic ground-state energies for Li, Be, $\text{Be}^+$, B, $\text{B}^+$ and $\text{C}^+$ are in agreement with the ECG results to within 0.2 mHa, as shown in Figure \ref{fig:atom-ECG}. For these systems, the ECG results are converged to essentially the exact ground-state energies in both the clamped-ion and nonadiabatic cases. The difference between our DMC ground-state and ECG reference is the fixed-node error present in our wave functions. We would expect the clamped ion results to be more accurate than the nonadiabatic results, since the nonadiabatic wave functions are inherently more difficult to construct.  However, for the systems in Figure \ref{fig:atom-ECG}, this difference in quality is less than 0.1 mHa. In the case of Be, $\text{Be}^+$, and B,  the nonadiabatic wave function is actually more accurate than the corresponding clamped-ion wave function.

No reference calculations exist for the heavier atoms N,O, and F. However, it is possible to apply finite-mass correction~\cite{Davidson_Atoms,Cencek_LiH} (i.e., divide by $1+m_e/M$, where $m_e$ is the mass of an electron and $M$ is the mass of the nucleus) to the best clamped-ion references to estimate the nonadiabatic references. The energies for N, O, and F obtained in this way are -54.5871, -75.0647 and -99.7310 Ha, respectively. For the ionized states, we obtain -54.0525, -74.5643 and -99.0900 Ha. 

\begin{figure}[h]
\centering
\includegraphics[scale=.37]{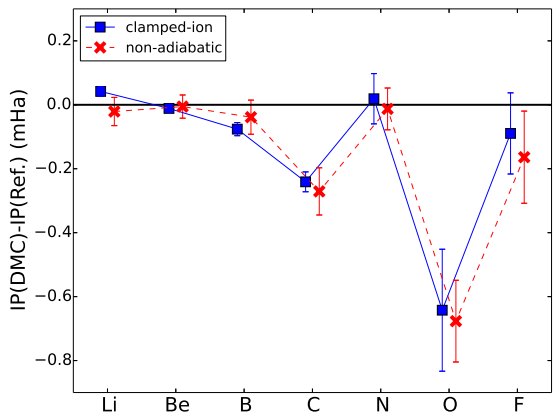}
\caption{Calculated ionization energies relative to reference data. The same reference is used for both clamped-ion and nonadiabatic results. The calculated energies are all within 1 mHa of the reference. \label{fig:ionization}}
\end{figure}

The ionization potentials are reported in Table \ref{tab:ionization} and shown in Figure \ref{fig:ionization}. For determining a set of nonadiabatic reference data, we subtract the spin-orbit and relativistic corrections (estimated by Klopper et. al.~\cite{Klopper_IP}) from the NIST experimental data.~\cite{NIST_Atoms} Ref.~\cite{Klopper_IP} is considered to have the most accurate ionization energies due to its usage of state-of-the-art quantum chemistry techniques shown to provide close agreement with experiment.
For the atoms considered in this work, ionization energies have previously been predicted to be independent of all nonadiabatic effects beyond the DBOC to within an accuracy of 0.1 mHa.~\cite{Klopper_IP} This prediction is based on calculations that are reported to be exact and agree to high accuracy with experiment. As shown in Figure \ref{fig:ionization}, the ionization potentials calculated with and without the Born-Oppenheimer approximation are all within 1 mHa of the reference energies. Further, the clamped-ion and nonadiabatic predictions for the ionization potentials are statistically indistinguishable for all systems studied, consistent with the previous study.~\cite{Klopper_IP}

\begin{figure}[h]
\includegraphics[scale=.4]{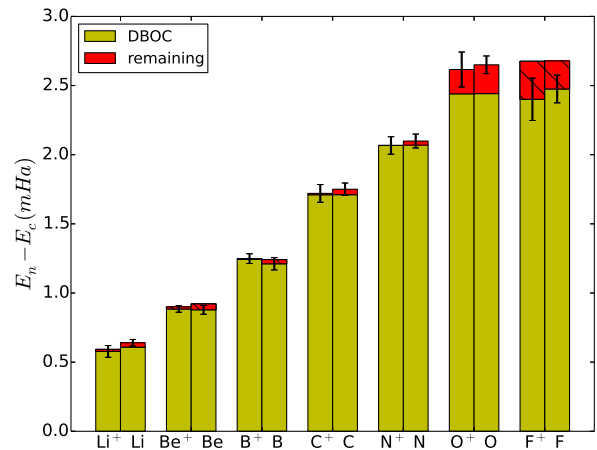}
\caption{The nonadiabatic contribution to ground-state energies of atoms and ions calculated with FN-DMC. The nonadiabatic contribution is partitioned into the DBOC and the remaining correction. A hatched bar indicates the contribution is negative. The numerical DBOC data is provided in Table \ref{tab:nad-ad-atoms}. \label{fig:atom-nad-ad}} 
\end{figure}

\begin{figure}[h]
\includegraphics[scale=.4]{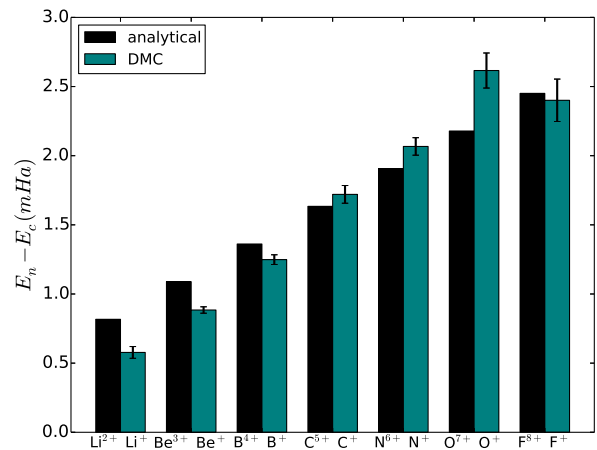}
\caption{The nonadiabatic contribution to ground-state energies of ions and their corresponding hydrogen-like atoms calculated with FN-DMC and analytically as shown in Eq. \ref{eq:analytical}. \label{fig:analytical}}
\end{figure}

\begin{table}[h]
\setlength{\extrarowheight}{1pt}
\caption{Nonadiabatic corrections for the ground-state energies of atoms and ions. $E_n$ and $E_c$ are the FN-DMC calculations of the nonadiabatic and clamped ground-state energies, respectively. The DBOC contribution is provided by Wim Klopper.~\cite{klop1} All energies are reported in units of mHa.\label{tab:nad-ad-atoms}}
\begin{tabular}{ccc|ccc}
\hline\hline
System & $E_n-E_c$&  DBOC     & System & $E_n-E_c$&  DBOC \\ \hline
Li$^+$ &   0.58(4) &  0.591970 & Li     &   0.64(2) &  0.608411 \\
Be$^+$ &   0.88(2) &  0.899706 & Be     &   0.88(3) &  0.920848 \\
B$^+$  &   1.25(4) &  1.242988 & B      &   1.21(5) &  1.241669 \\
C$^+$  &   1.72(6) &  1.710382 & C      &   1.75(5) &  1.710900 \\
N$^+$  &   2.07(6) &  2.066914 & N      &   2.10(8) &  2.069149 \\
O$^+$  &   2.6(1) &  2.440320  & O      &   2.6(2)&  2.441821 \\
F$^+$  &   2.4(2) &  2.675128  & F      &   2.5(1)&  2.678181 \\
\hline\hline
\end{tabular}
\end{table}

In Table \ref{tab:nad-ad-atoms} and Figure \ref{fig:atom-nad-ad}, we demonstrate the amount of nonadiabatic contribution to the ground-state energies in atoms and ions calculated as the difference between the nonadiabatic and clamped-ion ground-state energies. The amount of nonadiabatic contribution is always positive for these systems and mostly increases with atomic number. Using previous benchmark values for the DBOC, we can break down the nonadiabatic contribution of our system into a DBOC contribution and everything beyond the DBOC.~\cite{Klopper_DBOC,CFOUR,Harding} The DBOC is relatively insensitive to the level of theory. Figure \ref{fig:atom-nad-ad} indicates that in the atomic systems, the DBOC is the dominant contribution to the nonadiabatic energy, with the remaining amount being close to zero within error bars. The nonadiabatic energy is relatively constant between the neutral and cationic species. This observation suggests that the amount of nonadiabatic contribution is insensitive to the addition or removal of a valence electron. Physically, the valence electrons are farther from the nucleus than the core electrons, and thus are likely to be affected to a lesser degree by the delocalization of the nucleus. 

The nonadiabatic contributions in the cations can also be compared with those in their corresponding hydrogen-like atoms for a more in-depth analysis. The nonadiabatic contribution in a hydrogen-like atom can be obtained analytically. The result in Hartree atomic units is
\begin{align}
E_n-E_c=\frac{Z^2}{2}(1-\mu) \label{eq:analytical}
\end{align}
where $\mu=\frac{M}{M+1}$ is the reduced mass of the hydrogen-like atom and $M$ and $Z$ are the mass and atomic number of the nucleus, respectively. The increase in the nonadiabatic contribution with increasing $Z$ for hydrogen-like atoms reflects the stronger Coulombic attraction between the electron and the nucleus, which enhances the effects of the delocalization of the nucleus. An interesting case to consider is the transition from Li$^{2+}$ to Li. As shown in Figure \ref{fig:atom-nad-ad} and Figure \ref{fig:analytical}, the addition of a core electron to Li$^{2+}$ decreases the nonadiabatic contribution, while the addition of a valence electron has no further effect within our error bars. We also calculate the nonadiabatic contribution in Be$^{2+}$ to be $0.78(5)$ mHa, which is $0.29(5)$ mHa lower than the nonadiabatic contribution in Be$^{3+}$ and is closer to that in Be$^{+}$ of 0.88(2) mHa. Because the core electrons interact more strongly with the nucleus than do the valence electrons, the core electrons are affected more by the delocalization of the nucleus. Moreover, the addition of a second core electron decreases the nonadiabatic contribution for Li$^{2+}$ and Be$^{3+}$. 
We note that the nonadiabatic correction to the atomic ground-state energies of Eq. (\ref{eq:analytical}), which only holds for single electron systems, is roughly linear in Z, while the relativistic recoil correction~\cite{Merkt_H2} scales as $Z^4$. Therefore, the nonadiabatic effect is not seen experimentally,  as it is less significant than this relativistic effect.

\subsection{Hydrides}

\begin{table*}[t!]
\setlength{\extrarowheight}{1pt}
\begin{threeparttable}
\caption{Ground-state energies and atomization energies: fixed-node DMC results of this work for all first row hydrides with and without the Born-Oppenheimer approximation. The rows marked with bold \textbf{FN-DMC} are our nonadiabatic results. All atomization energies are estimated for 0K. $D_o$ includes zero-point energy contribution, while $D_e$ does not. Both total energies and dissociation energies are given in units of Hartree. \label{tab:atomization}}
\begin{tabular}
{
 l
 S[table-format=1.6]
 S[table-format=4.6]
 S[table-format=4.6]
 S[table-format=4.6]
 S[table-format=4.6]
 S[table-format=4.6]
 S[table-format=4.6]
}

\hline\hline
\multicolumn{1}{c}{Molecule} & 
\multicolumn{1}{c}{LiH$(^1\Sigma^+)$} &
\multicolumn{1}{c}{BeH$(^2\Sigma^+)$} &
\multicolumn{1}{c}{BH$(^1\Sigma^+)$} &
\multicolumn{1}{c}{CH$(^2\Pi)$} &
\multicolumn{1}{c}{OH$(^2\Pi)$} &
\multicolumn{1}{c}{HF$(^1\Sigma^+)$} \\ 
\hline
\multicolumn{1}{c}{} & 
\multicolumn{1}{c}{} &
\multicolumn{1}{c}{} &
\multicolumn{1}{c}{clamped-nuclei} &
\multicolumn{1}{c}{} &
\multicolumn{1}{c}{} &
\multicolumn{1}{c}{} \\
FN-DMC & \text{-}8.070518(7) & \text{-}15.24793(2) & \text{-}25.28867(3) & \text{-}38.4780(1) & \text{-}75.7356(1) & \text{-}100.4552(1) \\
$E_{\text{ref}}$ \tnote{a} & \text{-}8.0705473 & \text{-}15.2483(4) & \text{-}25.2893(2) & \text{-}38.4792(2) & \text{-}75.7382(2) & \text{-}100.4600(3) \\
\multicolumn{1}{c}{} & 
\multicolumn{1}{c}{} &
\multicolumn{1}{c}{} &
\multicolumn{1}{c}{nonadiabatic} &
\multicolumn{1}{c}{} &
\multicolumn{1}{c}{} &
\multicolumn{1}{c}{} \\
\textbf{FN-DMC} & \text{-}8.06624(3) & \text{-}15.24194(5) & \text{-}25.28128(9) & \text{-}38.4672(3) & \text{-}75.7245(5) & \text{-}100.4431(4) \\
ECG \cite{Bubin_LiH_noBO,Bubin_BeH_noBO,Bubin_BH_noBO} & \text{-}8.0664371(15) & \text{-}15.24203(10) & \text{-}25.2803(10) & N/A & N/A & N/A \\
\hline

\multicolumn{1}{c}{} & 
\multicolumn{1}{c}{} &
\multicolumn{1}{c}{} &
\multicolumn{1}{c}{clamped-nuclei} &
\multicolumn{1}{c}{} &
\multicolumn{1}{c}{} &
\multicolumn{1}{c}{} \\
$D_e$ (FN-DMC) & 0.09246(1) & 0.08062(2) & 0.13493(3) & 0.1335(1) & 0.1699(2) & 0.2234(1) \\
$D_e$ Feller \tnote{b} & 0.09262(5) & 0.0809(4) & 0.1354(2) & 0.1342(2) & 0.1709(2) & 0.2258(3) \\
\multicolumn{1}{c}{} & 
\multicolumn{1}{c}{} &
\multicolumn{1}{c}{} &
\multicolumn{1}{c}{nonadiabatic} &
\multicolumn{1}{c}{} &
\multicolumn{1}{c}{} &
\multicolumn{1}{c}{} \\
$D_o$ (\textbf{FN-DMC}) & 0.08910(4)  & 0.07578(6)  & 0.1290(1) & 0.1248(3) & 0.1617(5) & 0.2141(4) \\
$D_o$ Feller \tnote{c} & 0.08940(5) & 0.0761(4) & 0.1299(2) & 0.1276(2) & 0.1622(2) & 0.2166(3)\\
$D_o$ Exp. \cite{CCCBDB,HH} & 0.08874(38) & 0.07475(4) & 0.1281(37)\tnote{d} & 0.1275(5) & 0.1622(1) & 0.2158(3) \\
\hline\hline
\end{tabular}
\begin{tablenotes}
\item[a] For LiH, ECG provides the best reference energy.~\cite{Adamowicz_LiH} For the rest of the systems, we combined the best clamped-ion atomic references in Table \ref{tab:ionization} and thermochemistry estimates of $D_e$ in this table to produce the reference ground-state energies.
\item[b] Estimates for $D_e$ are calculated by subtracting the scalar relativistic, spin-orbit coupling and zero-point energy corrections from the reference $D_o$ in Table VI of Ref.~\cite{Feller_Corrections}.
\item[c] Here only the scalar relativistic and spin-orbit coupling corrections are subtracted.
\item[d] The atomization energy for BeH in Ref.~\cite{CCCBDB} disagrees with previous high-level theoretical benchmarks,~\cite{Feller_Corrections,Bubin_BeH_noBO} thus we use Ref.~\cite{HH} instead. For several of the systems, multiple experimental values are available in the literature.  We report experimental values that were aggregated in one single reference,~\cite{CCCBDB} except for BeH.~\cite{HH} 
\end{tablenotes}
\end{threeparttable}
\end{table*}
In Table \ref{tab:atomization}, we present our results on a series of molecular systems (hydrides). Finding accurate reference data for these systems to 0.1 mHa is not straightforward. We will use highly converged ECG data when available. Two ECG calculations have been performed in the clamped-nuclei limit for LiH~\cite{Cencek_LiH,Adamowicz_LiH} and we agree within 0.03 mHa with the more recent reference. For the rest of the systems, we combined the best clamped-ion atomic references in Table \ref{tab:ionization} and thermochemistry~\cite{Feller_Corrections} estimates of atomization energy $D_e$ in Table \ref{tab:atomization} to produce the reference ground-state energies. For BeH and BH, we are within 1 mHa of the reference values, and our energies are lower than the best available quantum chemistry results of -15.247846 Ha~\cite{Koput_BeH} and -25.287650 Ha~\cite{Miliordos_BH} for BeH and BH, respectively. 

\begin{figure}[h]
\centering
\includegraphics[scale=.37]{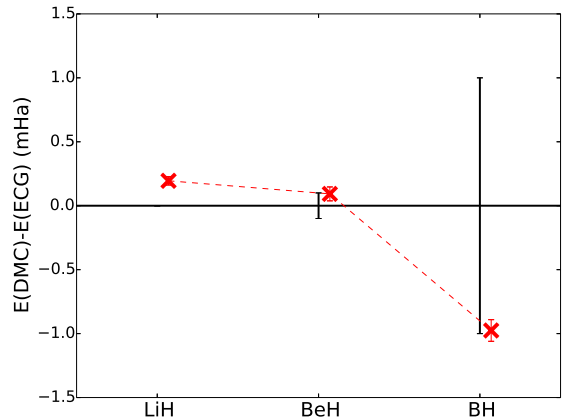}
\caption{The nonadiabatic FN-DMC ground-state energies of LiH, BeH and BH relative to ECG references. The error bars for the nonadiabatic ECG references are shown as thick dark lines, and the error bars for the FN-DMC calculations are comparable to the size of the symbols. \label{fig:dia-ECG}}
\end{figure}

Nonadiabatic ECG calculations only exist for the three smallest hydrides. Our results for LiH and BeH agree with the ECG references to within 0.2 mHa, as shown in Figure \ref{fig:dia-ECG}. The ECG reference for LiH is converged to the true ground-state energy beyond 0.1 mHa; thus, it is likely that our wave function has a fixed-node error of 0.2 mHa. For BeH, our result is within 0.1 mHa of the ECG reference and agrees within error bars. With BH being one of the largest ECG simulations performed, the DMC result is actually lower in energy, in this case by 1 mHa. The ECG error bar on BH is large, and it is not evident how close our result is to the true ground state, although extrapolating the ECG result with basis set size suggests we are within 1 mHa.~\cite{Bubin_BeH_noBO} For these nonadiabatic systems, we have the lowest variational result for BH, and the only simulated results of for CH, OH, and HF, to the best of our knowledge.

The atomization energies of the diatomic systems are reported in Table \ref{tab:atomization}. High-quality thermochemistry benchmarks are used for comparison.~\cite{Feller_Corrections} We take the reference energies from the last column of Table VI of Ref.~\cite{Feller_Corrections} and subtract the corrections in the $\Delta E_{SR}$ (scalar relativistic) and SO (spin-orbit coupling) columns for the comparison with our nonadiabatic energies. For the comparison with our clamped-nuclei results, we further subtract the DBOC and ZPE (zero-point energy) corrections. Corrections from spin-orbit coupling and relativistic effects are not used, as they are not included in our Hamiltonian. The atomization energies estimated in the clamped-nuclei limit agree within 1 mHa of the references for all but the largest molecule, HF. Within quantum Monte Carlo, it is generally more difficult to obtain an accurate nodal surface for a molecule than for an atom. As a result, our estimates for the clamped-nuclei atomization energies are lower than the references in all cases. A similar trend can be observed when comparing our nonadiabatic results with the references. For each molecule, the deviation from the reference is similar in the clamped-nuclei and nonadiabatic cases except for CH.

\begin{figure}[h]
\centering
\includegraphics[scale=.4]{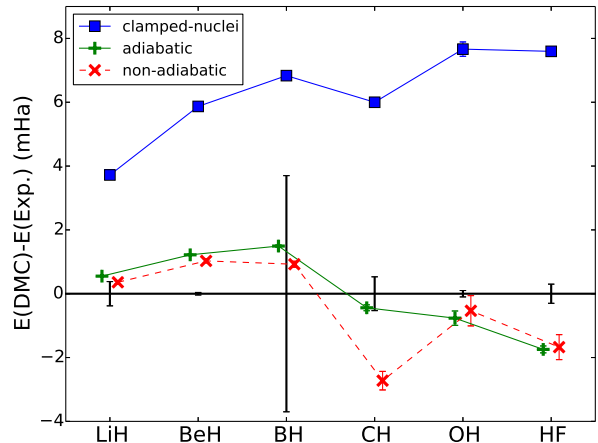}
\caption{Atomization energies of first row hydrides obtained with FN-DMC relative to experimental data. The adiabatic results are estimated by adding zero-point energies from Ref.~\cite{Feller_Corrections} to the clamped-nuclei results. \label{fig:atomization}}
\end{figure}

In Figure \ref{fig:atomization}, we compare both our clamped-nuclei and our nonadiabatic results to experimental data. We also provide adiabatic estimates by adding the zero-point energies calculated with coupled-cluster techniques in Ref.~\cite{Feller_Corrections} to our clamped-nuclei results. To calculate experimental atomization energies starting from the clamped-nuclei results, energetic corrections due to zero-point motion of the nuclei, nonadiabatic effects, spin-orbit coupling and relativistic effects should be included. For these highly adiabatic systems, the inclusion of zero-point motion alone is sufficient to bring our clamped-nuclei results to within 2 mHa of the experimental results. Except for the case of CH, the nonadiabatic results agree closely with their adiabatic counterparts and are closer to the experimental values, although for BH the experimental error bar is too large to provide a high-accuracy comparison. For CH, the experimental result suggests that our electron-ion wave function for this molecule has an unusually large fixed-node error.

\begin{figure}[h]
\includegraphics[scale=.4]{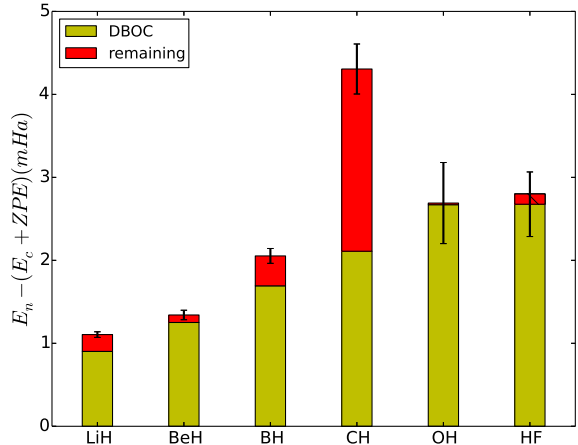}
\caption{The nonadiabatic contribution to the ground-state energies in hydrides calculated with FN-DMC. The adiabatic reference energies are calculated by adding zero-point energy contributions from Ref.~\cite{Feller_Corrections} to our clamped-nuclei results. The nonadiabatic contribution is partitioned into the DBOC and the remaining correction. A hatched bar indicates the contribution is negative. \label{fig:dia-nad-ad}}
\end{figure}

To estimate the nonadiabatic contribution to the ground-state energies for these hydrides, we calculate the difference between our nonadiabatic and adiabatic results, as shown in Figure \ref{fig:dia-nad-ad}. Similar to the atomic case, we break down the nonadiabatic energy of our system into a DBOC contribution and everything beyond the DBOC.~\cite{CFOUR,Harding,Feller_DBOC} The ZPE and DBOC contributions to this difference are listed in Table \ref{tab:nad-ad-diatomics}. We also calculate the nonadiabatic correction to the dissociation energies of the hydrides. 
For BeH, OH, and HF, the nonadiabatic contribution is almost entirely accounted for by the DBOC with the remaining correction being zero within error bars. For LiH, BH, and CH, the remaining amount of nonadiabatic contribution seems to be nonzero, and appears quite significant in CH. However, if the electron-ion wave function is significantly lower in quality than the electronic wave function for a given system, then the amount of nonadiabatic contribution will be overestimated. We also use the zero-point energies from Feller et. al.~\cite{Feller_Corrections} as corrections, which may introduce some additional uncertainty.  Regardless, our current predictions suggest that nonadiabatic effects in BH and CH are larger than in the other systems we considered.

For the LiH molecule, we also calculated the electron affinity for comparison to ECG results. We calculated the ground-state energy of LiH$^-$ to be $-8.08222(2)$~Ha for the case of clamped-nuclei. With nonadiabatic effects included, our result is  $-8.07811(3)$~Ha. Our nonadiabatic result is in good agreement with a previous ECG study,~\cite{Bubin_LiH-_noBO} which reported a value of $-8.07856887$~Ha. We report an electron affinity of $0.01187(4)$~Ha, which can be compared to the ECG prediction of $0.012132(2)$~Ha and agrees with the experimental value of $0.0126(4)$ Ha.~\cite{switch}

\begin{table}[h]
\setlength{\extrarowheight}{1pt}
\caption{Nonadiabatic corrections for the ground-state energies of diatomic molecules. $E_n$ and $E_c$ are the FN-DMC calculations of the nonadiabatic and clamped ground-state energies, respectively. The ZPE and DBOC contributions are provided by David Feller.~\cite{fell1} The nonadiabatic correction for the dissociation energy estimated with FM-DMC are included in the $\Delta D_o$ column. All energies are reported in units of mHa.\label{tab:nad-ad-diatomics}}
\begin{tabular}{ccccc}
\hline\hline
System & $E_n-E_c$ &   ZPE &      DBOC & $\Delta D_o$\\ \hline
LiH  &   4.28(3) &  3.17 &  0.902410 & -0.19(4) \\
BeH  &   5.99(6) &  4.65 &  1.251000 & -0.19(6) \\
BH   &   7.39(9) &  5.34 &  1.692559 & -0.6(1) \\
CH   &  10.8(3) &  6.44 &  2.109487  & -2.3(3)  \\
OH   &  11.1(5) &  8.43 &  2.670397  &  0.2(5)  \\
HF   &  12.0(4) &  9.34 &  2.799624  &  0.1(4)  \\
\hline\hline
\end{tabular}
\end{table}

\section{Conclusion}
We calculated the ground-state energies of first-row atoms and their corresponding ions and hydrides with and without the Born-Oppenheimer approximation. In addition, we examined the amount of nonadiabatic contribution to the ground-state energies of all systems studied and determined the amount to be up to a few mHa. In the case of CH, the nonadiabatic effects beyond the DBOC were unusually large, although we could not rule out the possibility that this discrepancy is due to the fixed-node error in our simulations. 
We found the ionization energies of the atoms to be independent of the Born-Oppenheimer approximation, consistent with a previous high-level quantum chemistry study.~\cite{Klopper_IP} In contrast, the atomization energies of the hydrides showed effects of nonadiabaticity, although they were generally much less than 1 mHa. This work obtained the first nonadiabatic QMC benchmark data for non-relativistic ground-state energies and obtained the lowest variational result for BH and the only results for CH, OH and HF, to the best of our knowledge.

In comparing to accurate benchmark results obtained with other methods, we have demonstrated the validity of our wave function ansatz, namely it does produce a high-quality electron-ion wave function. This technique also has the potential to solve interesting larger-scale problems due to its ease of implementation, as well as the polynomial scaling in computational time with respect to the number of electrons.

\section{Acknowledgment}
The authors would like to thank Mike Pak, Kurt Brorsen, Katharina Doblhoff-Dier and Brian Busemeyer for useful discussions. The authors would also like to thank Wim Klopper for providing the DBOC references for the atoms and ions and David Feller for providing the DBOC data for the hydrides. This work was supported by the U.S. Department of Energy (DOE) Grant No. DE-FG02-12ER46875 as part of the Scientific Discovery through Advanced Computing (SciDAC) program. NT and DC were supported by DOE DE-NA0001789. S.H.-S. acknowledges support by the National Science Foundation under CHE-13-61293. J.T.K. was supported through Predictive Theory and Modeling for Materials and Chemical Science program by the U. S. Department of Energy Office of Science, Basic Energy Sciences (BES). We used the Extreme Science and Engineering Discovery Environment (XSEDE), which is supported by the National Science Foundation Grant No. OCI-1053575 and resources of the Oak Ridge Leadership Computing Facility (OLCF) at the Oak Ridge National Laboratory, which is supported by the Office of Science of the U.S. Department of Energy under Contract No. DE-AC05-00OR22725.

\bibliography{ref}

\begin{thebibliography}{91}
\bibitem [{mas()}]{masses}%
  \BibitemOpen
  \href@noop {} {\ }\bibinfo {note} {All calculations are performed for the
  most abundant isotope. In units of electron mass, the isotope masses for Li,
  Be, B, C, N, O, F are taken to be 12782.4327, 16419.2608, 20214.7648 6,
  21862.7553, 25512.1484, 29141.0754, 34613.1200, respectively. The Li mass
  used for the LiH molecule is 12649.6690, which is slightly different from
  that used for the atomic Li simulations, but we do not expect this to affect
  our results within our statistical errors.}\BibitemShut {Stop}%
\bibitem [{Klo()}]{Klopper_DBOC}%
  \BibitemOpen
  \href@noop {} {\ }\bibinfo {note} {The DBOC values for the atoms and ions
  provided by Prof. Wim Klopper are calculated at the CCSD/d-aug-cc-pwCVQZ
  level using CFOUR.}\BibitemShut {Stop}%
\bibitem [{Fel()}]{Feller_DBOC}%
  \BibitemOpen
  \href@noop {} {\ }\bibinfo {note} {The DBOC references provided by Prof.
  David Feller are calculated at the CCSD(T)/aug-cc-pVTZ level using
  CFOUR.}\BibitemShut {Stop}%
\bibitem [{swi()}]{switch}%
  \BibitemOpen
  \href@noop {} {\ }\bibinfo {note} {We note that LiH ground state energies
  which we compare against are mislabeled in Ref.~\cite{Bubin_LiH-_noBO}, with
  $\text{LiH}^-$ and LiD being switched.}\BibitemShut {Stop}%
\end{thebibliography}
\end{document}